\documentclass[12pt]{iopart}

\usepackage{bm}
\usepackage{graphicx}
\usepackage{epsfig}
\usepackage{color}

\newcommand{\qav}[1]{\left\langle #1 \right\rangle}

\newcommand{\rem}[1]{}
\newcommand{\refe}[1]{(\ref{#1})}
\newcommand{\fige}[1]{Fig.~\ref{#1}}
\newcommand{\refE}[1]{Eq.~(\ref{#1})}
\newcommand{\beq}{\begin{equation}}
\newcommand{\eeq}{\end{equation}}
\newcommand{\beqa}{\begin{eqnarray}}
\newcommand{\eeqa}{\end{eqnarray}}

\begin{document}
\title[AC driven charge shuttle]
{Dynamics and Current Fluctuations in AC driven Charge Shuttle}

\author{F. Pistolesi$^1$ and Rosario Fazio$^2$}

\address{$^1$ Laboratoire de Physique et Mod\'elisation des Milieux
        Condens\'es, CNRS-UJF B.P. 166, F-38042 Grenoble, France}
\address{$^2$ International School for Advanced Studies (SISSA)
        via  Beirut 2-4,  34014 Trieste, Italy\\
    NEST-INFM $\&$ Scuola Normale Superiore
        Piazza dei Cavalieri 7, I-56126 Pisa, Italy}

\date{\today}

\begin{abstract}
The behavior of a charge shuttle under a pure AC field has been recently considered
theoretically and experimentally.
If the system presents an asymmetry in the tunneling amplitudes
the device acts as a nano-electromechanical rectifier, transforming a pure
AC voltage field into a direct curren.
In this paper we first review the model and the appearance of the rectifying effect
for bias voltages below the threshold of self-oscillation.
We discuss in some details the dynamics of the central island that, like
the current, presents strong dependence on the forcing AC field frequency.
In presence of both a constant and a small oscillating bias voltage we analyze the
transition from the static to self-oscillating solution.
We then consider current fluctuations (full counting statistics) for periodic motion
of the grain.
We explicitly evaluate the current noise numerically and we find that
it shows clear signatures of correlated transport at certain
locking frequencies.
In the adiabatic limit we obtain a simple expression for the full-counting statistics and
calculate explicitly the first four moments.
\end{abstract}

\submitto{}
\maketitle

\bibliographystyle{unsrt}

\section{Introduction}

The field of Nano-Electro-Mechanical (NEMS)
devices has been growing very rapidly in recent years
both experimentally and theoretically.
Important experimental results in this area include the use of a
Single Electron Transistor (SET) as
displacement sensor~\cite{knobel:2003} or the study of quantum transport through
suspended nanotubes~\cite{leroy:2004,sazanova:2004,sapmaz:2005}, oscillating
molecules~\cite{park:2000,smit:2002,kubatkin:2003,pasupathy:2005} and
islands~\cite{erbe:2001,scheible:2004}.
On the theoretical side, several
works~\cite{gorelik:1998,weiss:1999,boese:2001,armour:2002,braig:2003,sapmaz:2003,
novotny:2003,fedorets:2004,
armour:2004,pistolesiFCS:2004,novotny:2004,blanter:2004,chtchelkatchev:2005}
have highlighted various aspects of the role of Coulomb blockade in NEMS.

An exciting prototype example of mechanical assisted SET device is the charge shuttle
proposed in 1998 by Gorelik {\em et al.}~\cite{gorelik:1998}.
As shown in Ref.~\cite{gorelik:1998} a SET with an oscillating central island
can shuttle electrons between the two electrodes with a corresponding low noise
level~\cite{weiss:1999,pistolesiFCS:2004,novotny:2004}.
Although the experimental realization of the charge shuttle is difficult,
promising systems as  $C^{60}$ molecules in break
junctions~\cite{park:2000,pasupathy:2005} or silicon structures~\cite{erbe:2001,scheible:2004}
already exist.

Several interesting features in the dynamics of the shuttle emerge when it is driven
by an AC bias due to the interplay between the internal frequency of the oscillating
island and the external frequency~\cite{pistolesi:2005}.
The most striking is probably a rectification effect when the contact resistances to the
reservoirs are asymmetrical.
Moreover since the whole shuttling phenomenon is due to the non-linear dependence of
the electron transition rates on the position of the island one expects that the mechanical
motion can respond also at multiple harmonics both of the external and of the internal
resonating frequency.

In the present paper we expand our results on the AC-driven shuttle presented
in~\cite{pistolesi:2005} and discuss in some details the behavior of the average
current and its fluctuations together with a detailed analysis of the dynamics of
the island.
The paper is divided as follows.
In \Sref{model} we present the model for the AC forced shuttle.
In \Sref{current} we describe the dynamical behavior of the island and the
current dependence on the external frequency.
In \Sref{transition} the transition from a static to an oscillating solution is studied in presence
of a small AC field.
In \Sref{FCS} we derive an expression for the FCS in the adiabatic limit where we also
discuss the first four moments of current fluctuations.
\Sref{Conclusions} is devoted to the conclusions

\section{The model for the AC-driven charge shuttle}
\label{model}

%
%
\begin{figure}
    \centerline{
        \psfig{file=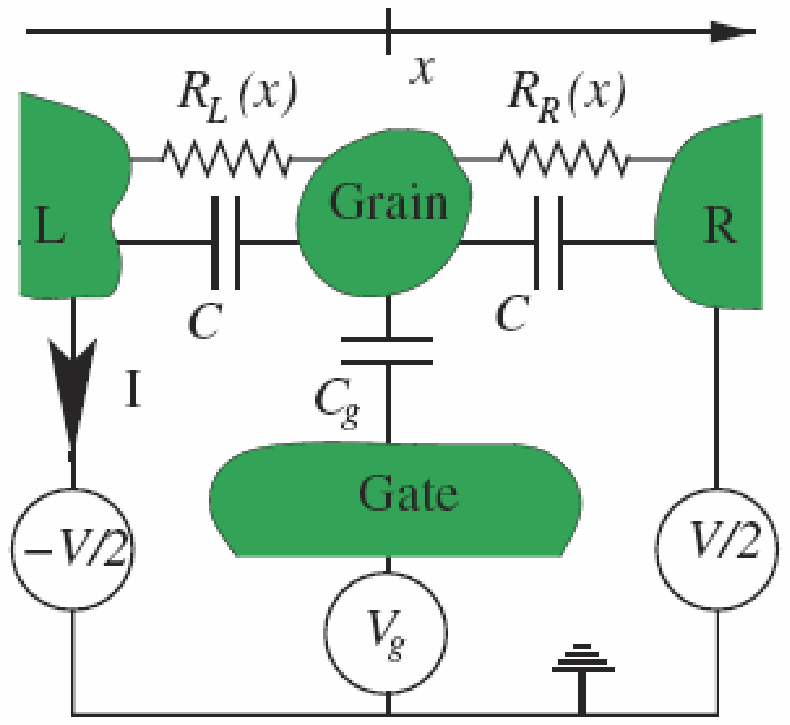,width=5cm,angle=-90}
        \psfig{file=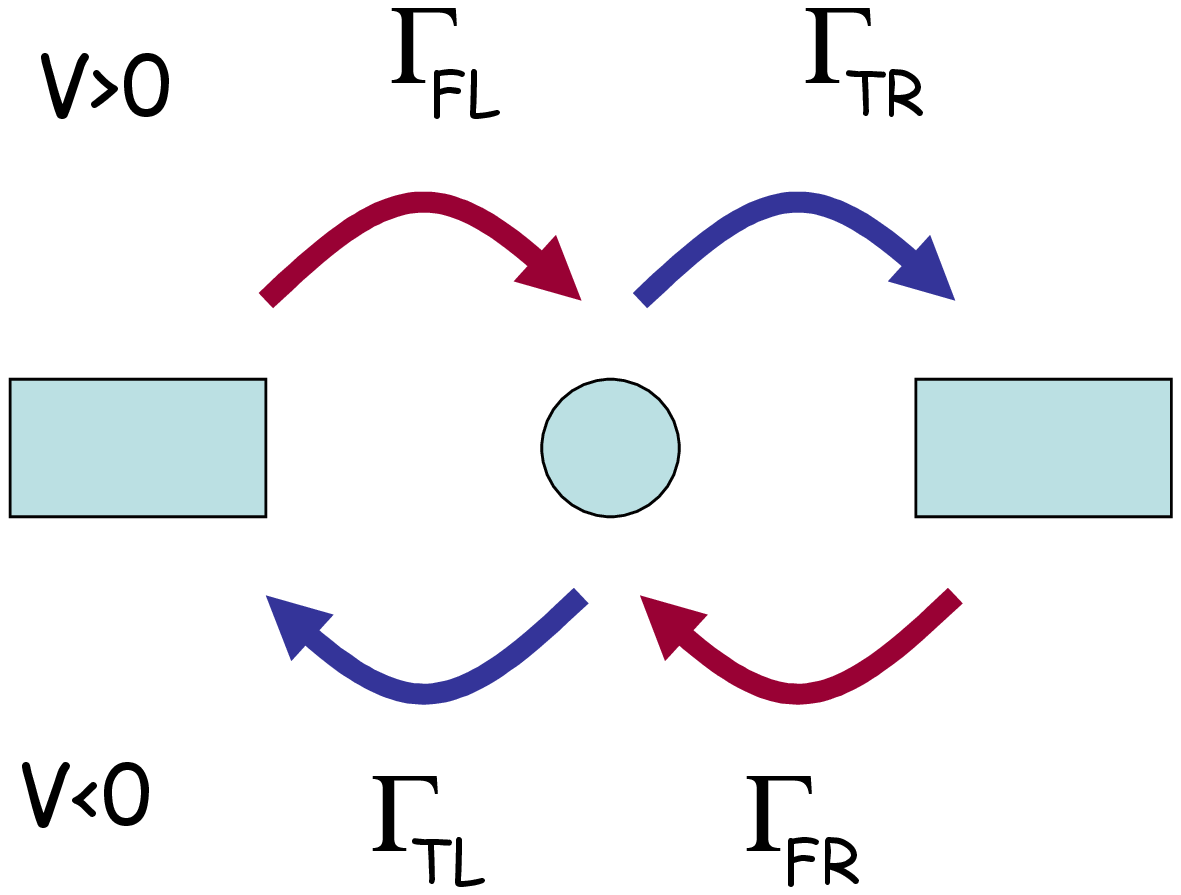,width=3cm,angle=-90}
        \psfig{file=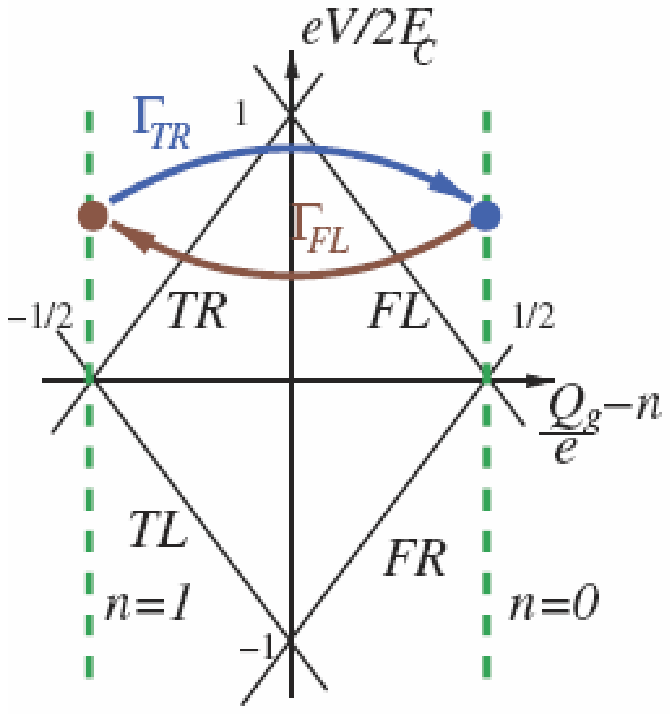,width=5cm,angle=-90}
    }
    \caption{Left panel: Electric scheme of a charge shuttle.
    Central Panel: Convention for the rates of charge transfer.
    Right panel: energy diagram for the SET. The diagonal lines indicate the
    thresholds for the vanishing of the four rates $\Gamma_{FL}$,
    $\Gamma_{FR}$,  $\Gamma_{TL}$, and  $\Gamma_{TR}$.
    The two dots indicate the state of the system during shuttling at
    positive voltage $V$.}
\label{fig1}
\end{figure}
%
%

The charge shuttle, shown in \Fref{fig1} left panel, consists in a
SET where the central island can oscillate between the two
leads~\cite{gorelik:1998} and  subjected to an elastic recoil force, a damping
force due to the dissipative medium, and an electric force due to the
applied bias.
The island is connected to the left and right leads through tunnel
junctions characterized by resistances $R_L(x)=R_L(0) e^{x/\lambda}$ and
$R_R(x)=R_R(0) e^{-x/\lambda}$. Here $\lambda$ is the tunnelling length
of the order of one nanometer, and $x$ is the displacement from the equilibrium
position in absence of external drive.
The island is coupled to the leads and the gate through two
capacitances $C_g$ and $C$.
While it is crucial to keep into account the (exponential) dependence of the
tunneling resistance on the position of the island, the dependence on $x$
of the capacitances is much weaker and it can be ignored.
The system is symmetrically biased at a voltage $V(t)$
($V_R=-V_L=V/2$), the charging energy is $E_C=e^2/2C_\Sigma$ where
$e$ is the electron charge, $C_\Sigma=C_g+2C$, and the gate charge $Q_g$
is $C_g V_g$.
We will consider the case of low temperatures ($k_B T \ll E_C$),
charge degeneracy ($Q_g = e/2$), and voltages $|V|<E_C/e$.
In this regime the grain can accommodate only $n=0$ or $1$ additional
electrons (see right panel of \Fref{fig1}).
We also assume that typical driving frequencies $\omega$ are small
compared to $k_B T/\hbar \ll E_C/\hbar$.
This condition on the frequency is rather weak (typically $E_C/\hbar \sim
10$ THz) but it allows a simple description of the tunnelling in terms of time
dependent rates.

In the simplest approximation the dynamics of the central island can be derived from
the Newton equation~\cite{gorelik:1998}
\beq
\ddot x(t) = -\omega_o^2\, x(t) - \eta \omega_o\, \dot x(t) + {eV(t)\over mL}\, n(t)
\quad.
        \label{xeq}
\eeq
together with a stochastic equation governing the charge $-en(t)$ on the island.
In Eq.(\ref{xeq})  $m$ is the mass of the grain, $\omega_o$ is the oscillator
eigenfrequency, $\eta \omega_o$ is a damping coefficient ($\eta\ll 1$),
and $L$ is the distance between the two leads.
In the regime of incoherent transport the (stochastic) evolution of
the charge $-en(t)$ is governed by the following four
rates~\cite{averin91}:
\beqa
    \Gamma_{FL} &=& |eV(t)/4E_C| \, \Gamma_L(x) \, \Theta(V)
    \label{n1a}
    \\
    \Gamma_{FR} &=&  |eV(t)/4E_C| \, \Gamma_R(x) \, \Theta(-V)
    \label{n1b}
    \\
    \Gamma_{TL} &=&  |eV(t)/4E_C| \, \Gamma_L(x)\, \Theta(-V)
    \label{n0a}
    \\
    \Gamma_{TR} &=&  |eV(t)/4E_C| \, \Gamma_R(x) \, \Theta(V)
    \label{n0b}
\eeqa
where Eqs.~(\ref{n1a}-\ref{n1b}) refers to $n=1\rightarrow0$ transitions while
Eqs.~(\ref{n0a}-\ref{n0b}) to $n=0 \rightarrow 1$ transitions.
Here $FL$, $FR$, $TL$, and $TR$ stands for From/To and Left/Right,
indicating the direction for the electron tunneling associated to the
corresponding $\Gamma$ as shown in the central panel of \Fref{fig1}.
$\Gamma_{L/R}(x)=[R_{L/R}(x)C]^{-1}$ and $\Theta(t)$ is the Heaviside
function.
The current $I$ is then determined by counting the net number of
electrons that have passed through the shuttle during a given time $t$.

\section{Current and mechanical response to the AC field}
\label{current}

\begin{figure}
\centerline{
\psfig{file=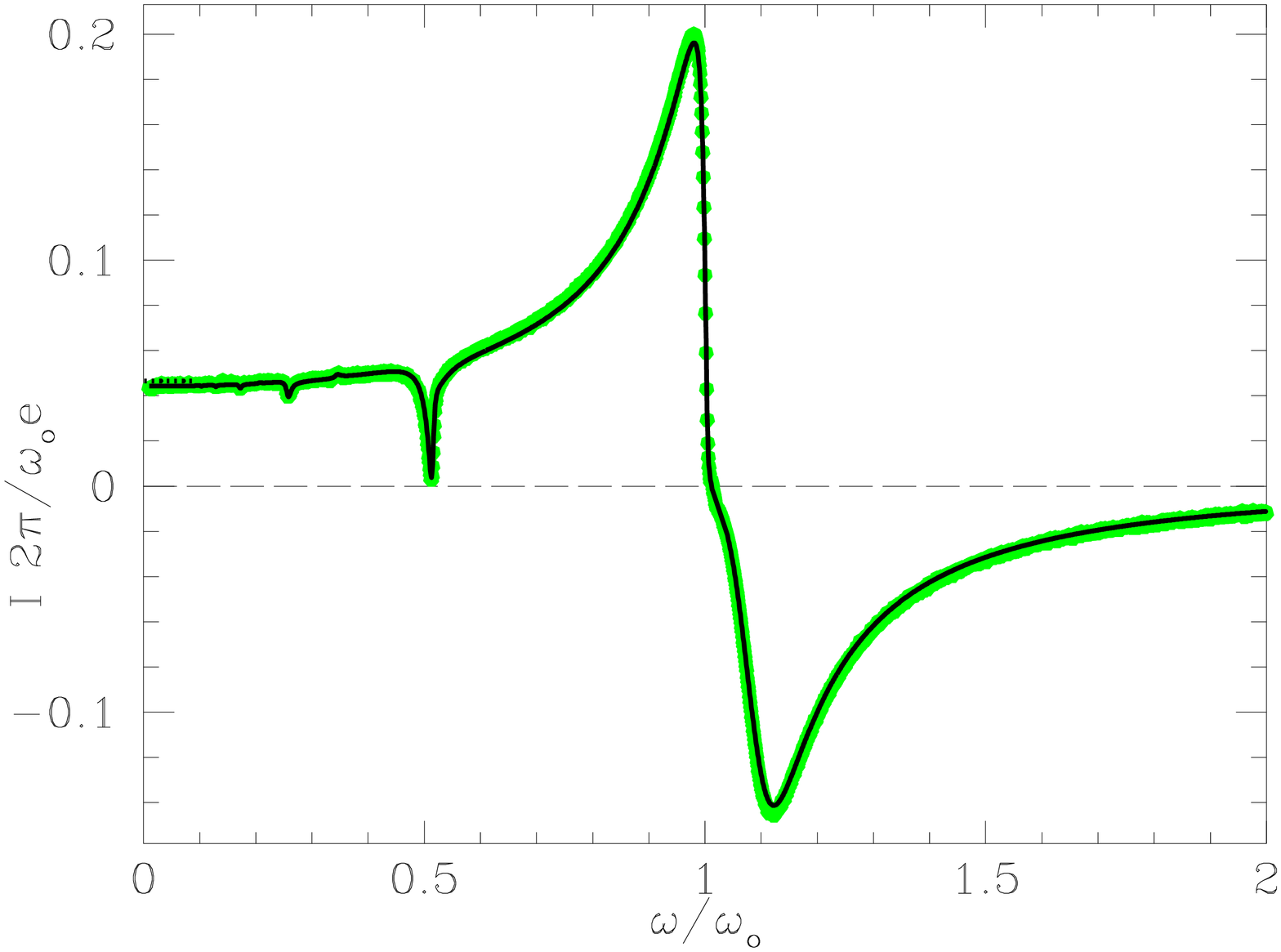,angle=-90,width=7cm}
\hskip.5cm
\psfig{file=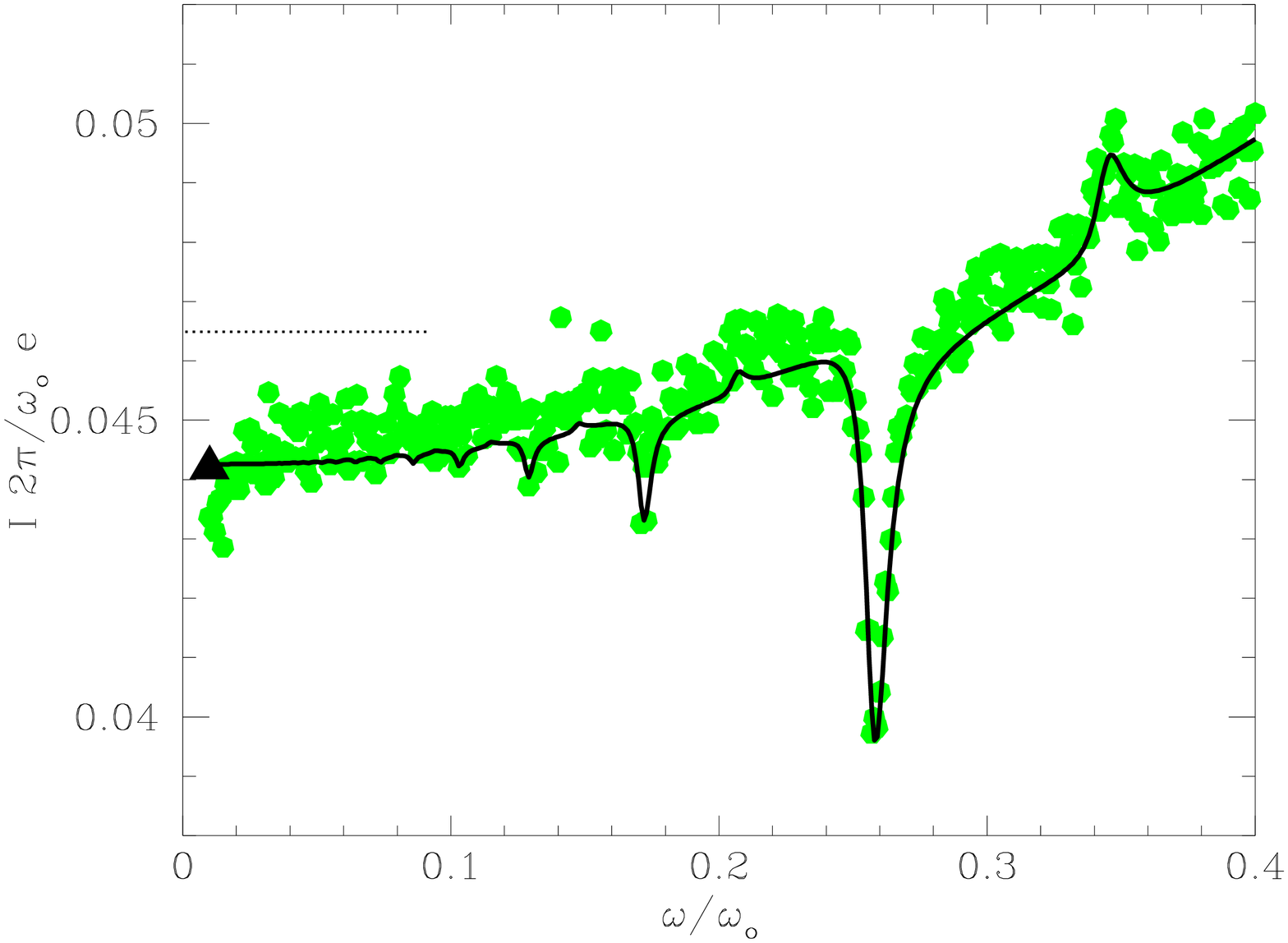,angle=-90,width=7cm}
}
\caption{Current as a function of the frequency for $\epsilon=0.5$,
  $\eta=0.05$, $\Gamma/\omega_o=1$, and $R_R/R_L=10$. The result of the
  simulation of the stochastic dynamics (green points) is compared with the
  approximate $I_a$ (full line). In the small frequency region, enlarged
  in the right panel, several resonances at fractional values of $\omega_o$ appear.
  We also show (red line) the analytical result from~\refE{adiabatic}
  in the adiabatic limit. The triangular dot indicates the numerical solution
  of the adiabatic equations.
}
\label{fig2}
\end{figure}

The problem can be studied numerically by simulating the
stochastic process governed by the four rates Eqs.(\ref{n1a}-\ref{n0b}) and by the
deterministic evolution~\refE{xeq}.
We used a simple technique that consists in solving analytically Newton equations between
two tunneling events.
One divides the trajectory into small intervals where the probability of occurrence
of two tunneling events is much smaller than one, and uses the rates to generate events
with the appropriate statistics.
In the following we present results for typically $10^6$ tunneling events for each plotted point.
After a transient time the system reaches a stationary behavior.
In \Fref{fig2} the stationary DC current is shown as a function of the frequency of the external bias.
The rich frequency landscape structure is generic. We observed a qualitatively
identical behavior in a wide range of parameters for which the dissipation is
sufficiently large to prevent self oscillation at constant voltage.
The first conclusion one can draw is the existence of a rectification
effect.
The effect is of truly nanoelectromechanical origin and most probably
it has been already observed by Scheible and Blick~\cite{scheible:2004}.
They considered a nanopilar forced by an AC field, and they observed a rectification effects
in the MHz range.
The structure of the main resonance strongly resembles the prediction shown in \Fref{fig2}, with a
change of sign of the current at $\omega_o$.
The theory presented strictly does not apply to the experiment of Ref.~\cite{scheible:2004}
which was performed at room temperature, nevertheless a qualitative comparison seems to indicate
that the main features are present
even when the Coulomb blockade is not completely established ($E_C\approx 80$ K in
that experiment).

A second important point to emphasize in \Fref{fig2} is the presence of a rich structure
at low frequency.
This is a consequence of the non-linearity of the problem, since the dips and the peaks appear at
rational values of the ratio $\omega/\omega_o$ [cf. right panel of \Fref{fig2}].
In order to understand better the whole response it is useful to
exploit an analytical description of the shuttle within the approximation scheme
proposed in Ref.~\cite{gorelik:1998} which we discuss below.

%
%
%
%

\subsection{Mean field approximation}

If the electric force is much smaller than the mechanical one,
$\epsilon = e V_o/(\omega_o^2 m \lambda L)\ll 1$,
one can average the force generated by the stochastic
variable $n(t)$ over many periods of oscillation.
One should in fact realize that what matters are the
hopping events.
If the charge on the island does not vary the electric force
is constant and thus no energy can be pumped on the system.
It is thus the charge fluctuations that induce the changes in the dynamics.
In order to have a sizable effect at $\epsilon \ll 1 $ we need to take into
account many hopping events. This allows to neglect the fluctuations
with respect to the average, even if the time scales of the hopping
and of the oscillations are of the same order.
Formally this consists in substituting into \refE{xeq} its
statistical mean: $\qav{n(t)}=\sum _n n P_n(t)$ where $P_n(t)$ is the
probability that the occupation number of the island is $n$
(in our case only $n=0$ and $1$ are allowed values).

The charge dynamics in the central island is then determined by the
the coupled set of equation which include the dynamical equation for the position
of the island together  with the master equation~\cite{averin91} for the
probability vector  $|P\rangle = \{P_0, P_1 \}$.
\begin{eqnarray}
    &\ddot x(t) &= -\omega_o^2\, x(t) - \eta \omega_o\, \dot x(t) + {eV(t)\over mL}\, P_1(t)
    \label{xeqa}
    \\
    & \partial_t |P(t)\rangle & = -\hat \Gamma(t) |P(t)\rangle
\label{MasterEq}
\end{eqnarray}
The matrix $\hat \Gamma(t)$ is given by
\beq
    \hat \Gamma =
    \left(
    \begin{array}{cc}
    \Gamma_{FL}+\Gamma_{FR} & -\Gamma_{TR}-\Gamma_{TL}
        \\
    -\Gamma_{FL}-\Gamma_{FR} & +\Gamma_{TR}+\Gamma_{TL}
    \end{array}
    \right)
    \,.
    \label{Rates}
\eeq
The probability is conserved, {\em i.e.} $P_0+P_1=1$.
The rates  depend on time through the extenal voltage
$V(t)$ and through the position of the island $x(t)$.
The instantaneous (average) current through the structure is
\beq
    I_a(t)/e
    =
    P_0(t)\,\Gamma_{FL}(t)-P_1(t)\, \Gamma_{TL}(t)
    \quad,
    \label{Ileft}
\eeq
where the subscript $a$ in the current indicates that the fluctuations of
the force acting on the shuttle, due to the discrete nature of the charge
tunneling, are neglected.
As shown in Ref.~\cite{gorelik:1998} the shuttle instability, at constant
bias, is controlled by the ratio $\epsilon/\eta$.
It is thus possible to assume that both $\epsilon$ and $\eta$ are
small, with their ratio arbitrary.

We are now ready to discuss the dependence of the current on the
external frequency.
The most prominent structure, observed also in the experiments of
Ref.~\cite{scheible:2004}, is present  at $\omega\approx
\omega_o$, and it corresponds to the main mechanical resonance.
The current changes sign across the resonance.
The shape of the resonance can be explained as an effect of the phase relation between
the driving voltage and the displacement of the grain.
A simple way to prove this fact is to solve \refE{MasterEq} for a given
function $x(t)= A \sin(\omega t-\phi)$ therefore neglecting higher harmonics.
The crucial parameter is then the phase $\phi$ and its frequency
dependence.
We found that if we substitute the usual relation for
a damped harmonic oscillator:
$\phi(\omega) = \arctan[ \omega \omega_o\eta /(\omega_o^2-\omega)]$
we could reproduce qualitatively the behavior of \fige{fig2} near
the resonance, and in particular the change of sign.
We verified from the numerical simulations that $\phi(\omega)$
around the main resonance agrees very well with the simple expression
of the damped harmonic oscillator given above.

%
%
\begin{figure}
\centerline{
\psfig{file=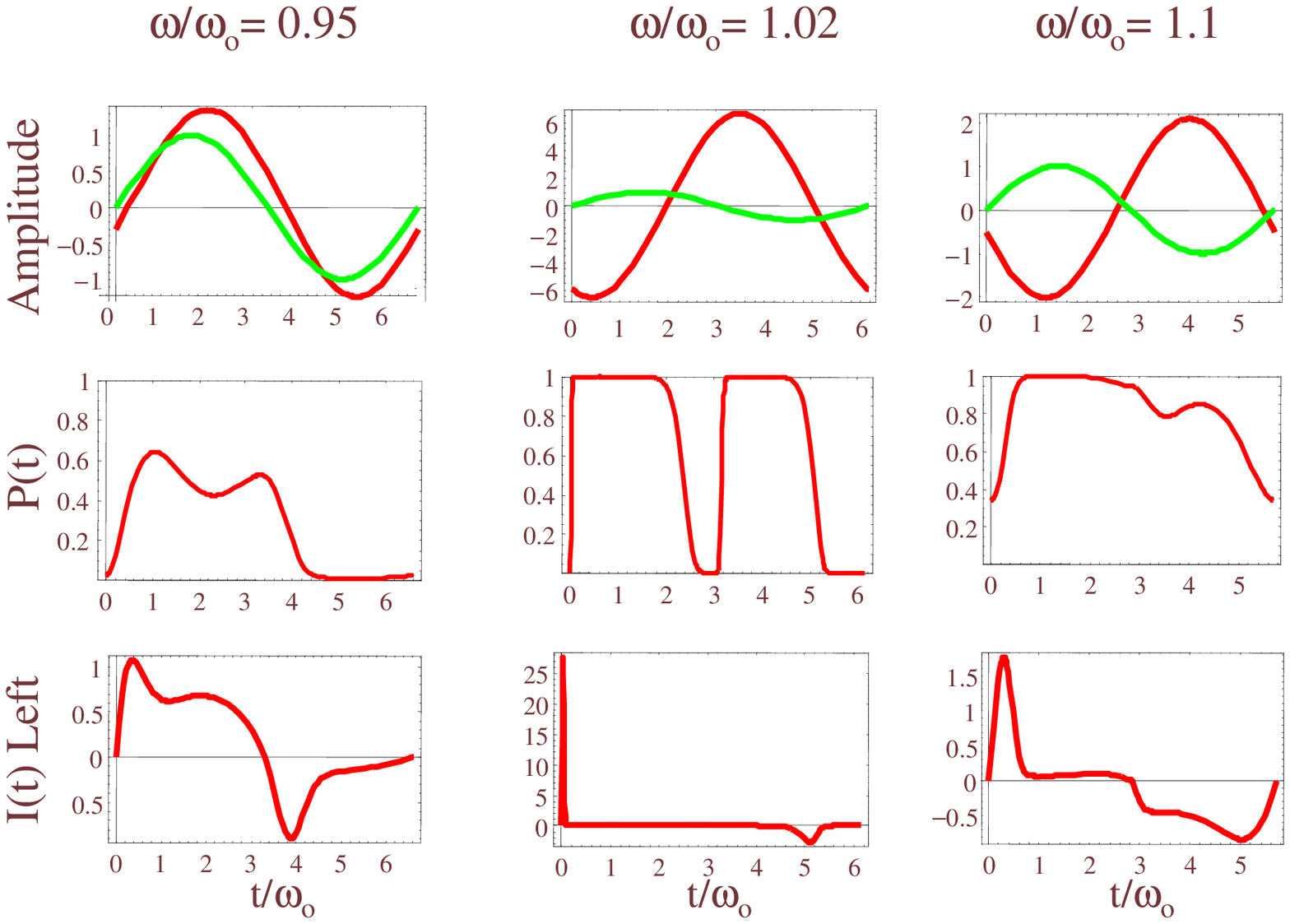,angle=-90,width=10cm}
}
\caption{
Behavior of  the position of the grain $x(t)/\lambda$ (red line, top row)
compared to the forcing field (green line, top row).
Probability $P_1(t)$ (central row) and current crossing the left junction (bottom row).
The three sets of plots refers to the positive peak, to the vanishing value and
to the negative peak in \Fref{fig2} near the main resonance.
}
\label{figRes}
\end{figure}
%
%

%
More insight can be gained by the solution of the coupled set of
Eqs.\refe{xeqa} and \refe{MasterEq} with periodic boundary conditions in order
to obtain the stationary behavior.
The results for three cases are shown in \Fref{figRes}.
The landscape near $\omega_o$ of the rectified current can be understood
by means of a following argument.
The total resistance of the structure, $R_T(x)=R_L(x)+R_R(x)$, has a minimum
for positive $x$ due to the asymmetry of the barriers.
Then, even if the transport is not given by Ohm's law, the fact that $R_T(x)$ has
a minimum shifted from the symmetric point allows larger currents to flow
when the shuttle is near the right lead.
For $\phi \approx 0$ the voltage is positive when the shuttle is on
the right, while the opposite is true for $\phi \approx \pi$.
The phase shift crosses sharply from 0 to $\pi$ when $\omega$
is swept through the resonance.
This explain while on the left of the resonance the rectified current
is positive and on the right it is negative.
The maximum is due to a combination of the fast frequency dependence of the
phase and of $amplitude$, that display the usual lorenzian maximum at resonance.
Between these two values the current has to vanish for
a value of the frequency that is very near $\omega_o$.

\begin{figure}
\centerline{\psfig{file=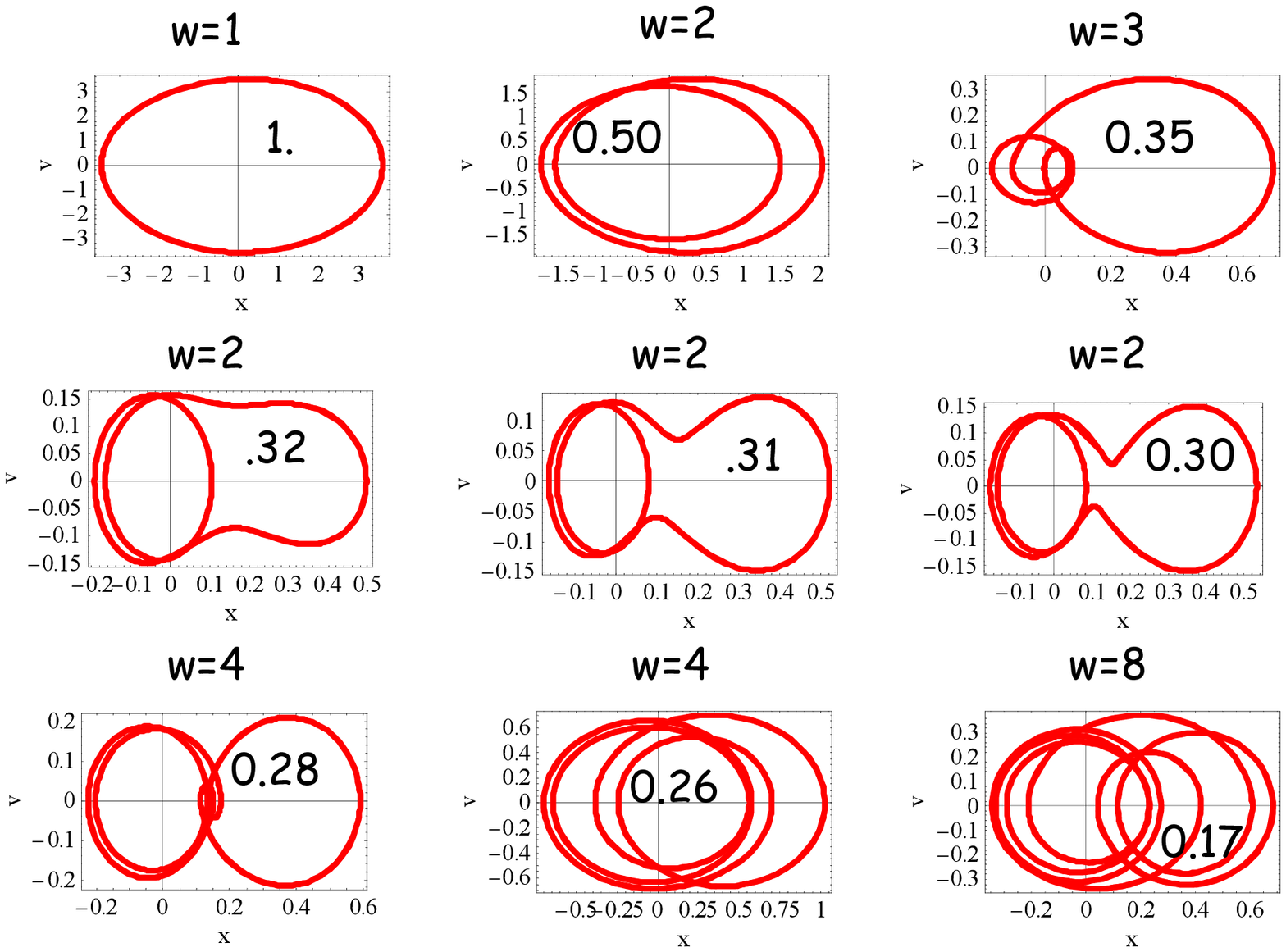,angle=-90,width=12cm}}
\caption{
Behavior of the shuttle in the phase space for different
winding numbers ($w$ is indicated in the figures).
The orizontal axis is the position in units of $\lambda$ and the vertical axis is
the velocity in units of $\lambda \omega_o$.
One can see for instance how the transition from $w=2$ to $w=4$ can
be performed without passing through the $w=3$ state.
}
\label{figW}
\end{figure}

From the right panel of \fige{fig2}, it is clear that additional structures appear
for $\omega \approx \omega_o/\nu$ with $\nu=2, 3, \dots $.
Our numerical simulations indicate that, except for
the fundamental frequency, even $\nu$ are favorite with respect to odd
ones.
As we already anticipated, the motion of the shuttle and the
oscillating source become synchronized at commensurate frequencies
whose ratio is $\mu/\nu$ (with $\mu$ positive integer).
This ratio, known also as the winding number,
can be defined as
\beq
   w =
   \lim_{t\rightarrow \infty}
   {\theta(t)/\omega t}
   \quad,
    \label{wdef}
\eeq
where $\theta(t)$ is the accumulated angle of rotation of
the representative point $(\dot x,\ddot x)$ in the
phase space.
In practice $\theta(t)/2\pi$ gives the number of oscillations
performed by the shuttle during the time $t$.
It is however important to choose as a representative point the velocity $\dot x$ and
the acceleration $\ddot x$ instead of the position $x$ and the
velocity $\dot x$.
In this way the angle really gives the number of "oscillations" performed by
the shuttle.
At large value of $w$ the structure can be quite complex as it can be seen from
\Fref{figW} and the above definition succeeds in giving the right number of closed
cycles in each case.
Again these plots have been obtained in the mean field approximation.
At the main resonances for $\omega/\omega_o\approx 1/\nu$, the shuttle performs
$\nu$ nearly harmonic oscillations per period of the forcing field.
Since different $w$ implies different topology in the phase space, it is
interesting to follow the transition from one winding number to an other.
In particular the transition from $w=2$ to $w=4$ is shown in \Fref{figW}.
This is performed without passing through the $w=3$ state, that appears actually
as a small plateau in the middle of the $w=2$ plateau.

%
%
%
%

\begin{figure}
\centerline{
\psfig{file=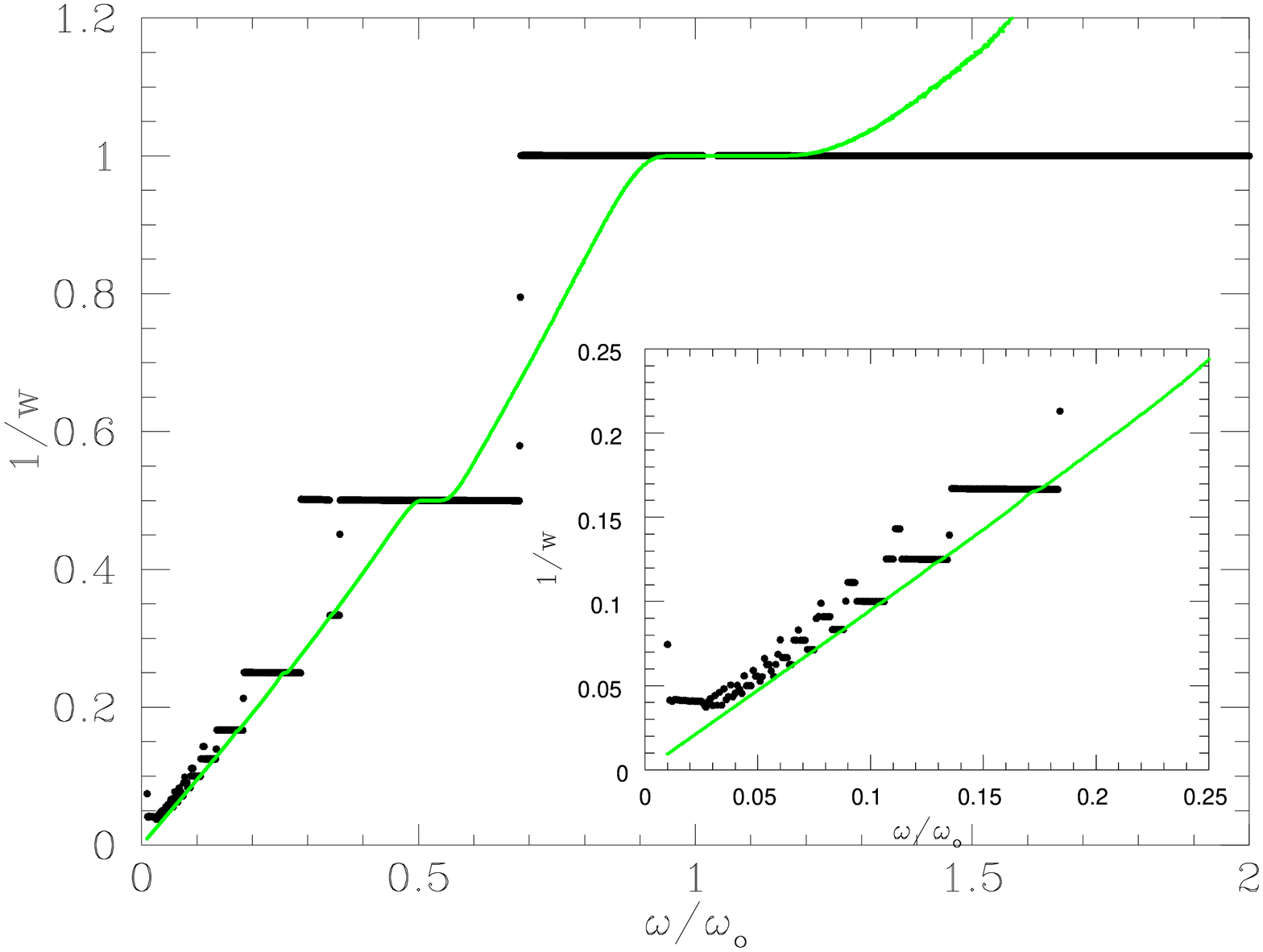,angle=-90,width=7cm}
\hskip.5cm
\psfig{file=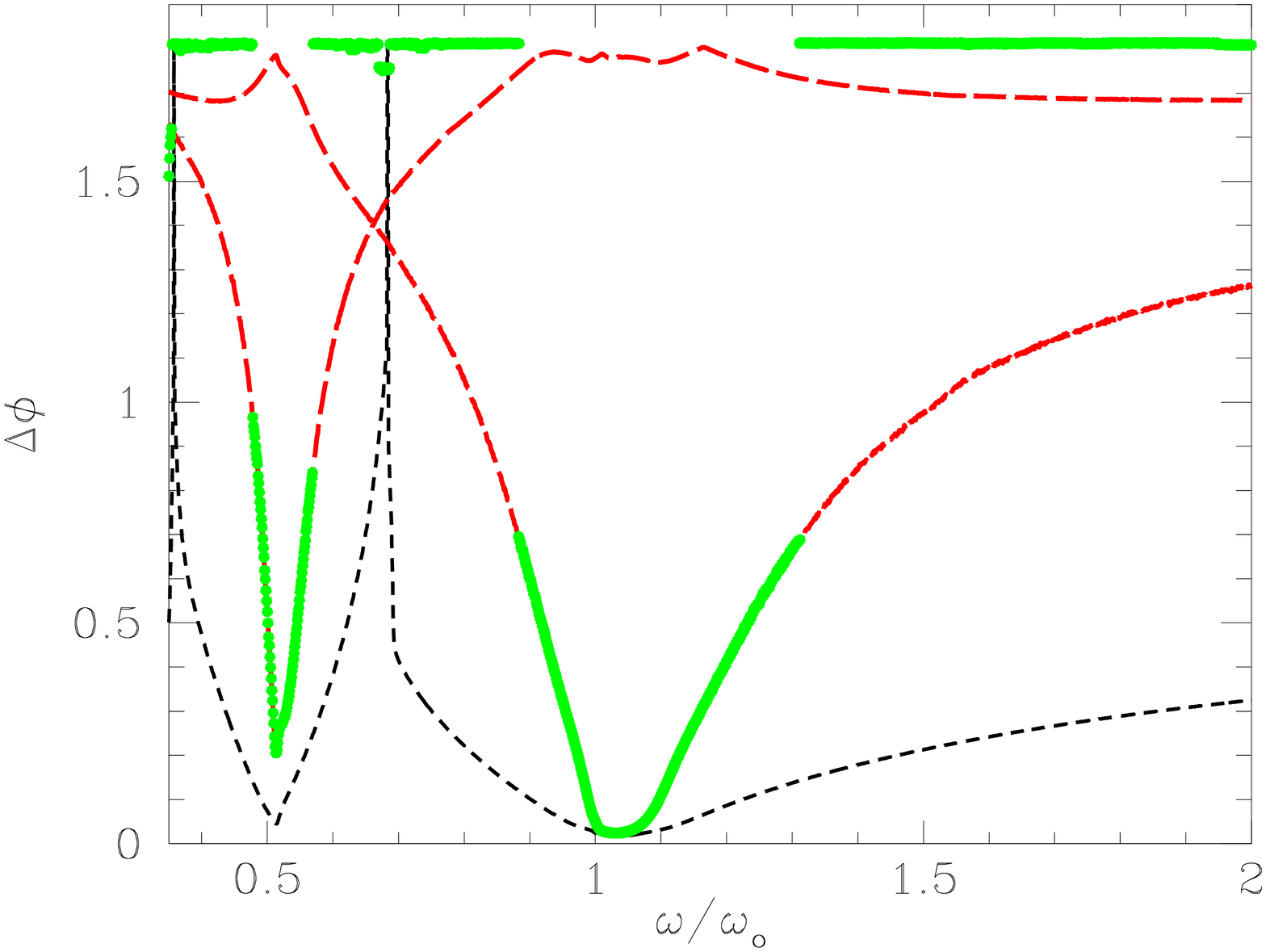,angle=-90,width=7cm}
}
    \caption{
Left panel shows the calculated winding number obtained either by the
stochastic simulation (green points) and the average approximation of
\refE{MasterEq} (black points). The inset enlarges the region near the origin.
Bottom panel shows $\Delta \phi$ obtained with different methods:
({\em i}) average approximation (full line),
({\em ii}) stochastic simulation with $w$ from \refE{wdef} (green points),
({\em iii}) stochastic simulation with $w$ set equal to 1 (red dashed line)  or
({\em iv}) with $w=2$ (blue dashed line).
 Parameters are the same as in \fige{fig2}.
}
\label{fig3}
\end{figure}

%
When the system is frequency locked at a winding number $w$, it is
possible to define the phase shift $\phi(t) = \theta(t)-w \, \omega
t$.
For an accurate analysis, the value used for $w$ in the definition
of $\phi(t)$ must be the best $\mu/\nu$ approximation to $w$, since even
corrections of the order of $10^{-3}$ to the exact $\mu/\nu$ results
would accumulate in the calculation of $\phi$, and give an arbitrary
shift of the order of $\pi$.
After a transient time for perfect locking, if the oscillation is
perfectly harmonic, $\phi$ should no more depend on $t$.
Thus an additional important quantity to analyze is the phase shift variance
$ (\Delta \phi)^2 = \qav{\phi^2}-\qav{\phi}^2$.
This is calculated by sampling $20$ points per cycle over $10^3$ cycles.
For a given $w$ the smaller is the value of $\Delta\phi$, the better the system
locks to that external frequency.
A zero value of the $\Delta\phi$ indicates that the oscillation is also
harmonic.
A periodic, but non-harmonic, oscillation
will induce a fluctuation of the phase in time, but
the fluctuations are correlated and $\Delta\phi$ will be smaller than the
completely uncorrelated result.
The numerical results for $w$ and $\Delta \phi$ are
shown in \fige{fig3}.
%
%
%
%
%
%
%
It shows the dependence of the
winding number as a function of the external frequency (left panel)
together with the analysis of $\Delta \phi$ (right panel)
calculated from the stochastic simulation, green points, and from the average
approximation, black line.
The locking at rational winding numbers is confirmed by the presence of
plateaux of decreasing width.
As expected the most stable plateau is at $w=1$: the system locks
very well at this frequency.
We find that this holds up to very high frequency in the
average approximation, where $w=1$ seems the only possible winding number.
The stochastic simulation would indicate instead that for
$\omega/\omega_o >  1.3$
the locking with $w=1$, as defined from \refE{wdef}, is no more established.
We verified that this upper limit to the $w=1$ plateaux depends essentially linearly
on the asymmetry $R_R/R_L$.
Quite generally plateaux size increases by increasing the asymmetry.
However, by studying $\Delta \phi$ for $w=1$ in the whole frequency
range, one actually obtains that a correlation is always present
({\em i.e.} $\Delta\phi < \pi/\sqrt{3}\approx 1.81$)
even when \refE{wdef} gives a value of $w$ different from one.
The stochastic fluctuations thus unlock the shuttle globally, but not locally.
Looking at the presence of local locking at other winding numbers
we find that for instance $w=2$ is clearly present for
$\omega>1$ and reversely $w=1$ is present around $\omega=1/2$
(see red dashed lines in the right panel of \fige{fig3}).
For global locking, only one phase variance is minimal.
It corresponds to the ``dominant'' winding number.
The presence of correlations of other winding numbers
may indicate partial locking at these winding numbers
(as in the region $0.6 \sim < \omega < \sim 0.9 $ for $w=1$ and $2$)
or the contribution of higher armonics of $x(t)$
(as in the region $\omega>1$).

%
%
%

\section{Instability in presence of the AC-drive}
\label{transition}

In the previous section we have seen the response of a shuttle to a pure
AC field for sub critical values of $\epsilon/\eta$, {\em i.e.} when the shuttle
instability is not present for static voltage.
In this section we describe the transition from the stable to the unstable
state in presence of a small forcing AC field.
When the instability develops continuously from the static solution
(soft excitation) one can study analytically the effect of an AC field
by expanding around the static solution.
We begin by solving the evolution equation in Fourier series:
\beq
    x(t)/\lambda = \sum_{\nu=1}^{\infty}\left[ A_\nu \sin(\nu \omega t) + B_\nu \cos(\nu \omega t)\right]
    + B_0/2
\eeq
The electric force for a static voltage can also be written as a Fourier series as follows:
\beq
    \sum_n  P_n(t) n = \sum_{\nu=1}^{\infty}\left[ C_\nu \sin(\nu \omega t) + D_\nu \cos(\nu \omega t)\right]
    + D_0/2
    \label{electricforce}
\eeq
where
$C_\nu=(\omega/\pi) \int_0^{2 \pi/\omega} \sin(\nu \omega t) \, \sum_n P_n(t)n$,
and $D_\nu$ has an identical expression with the cosine substituting the sine.
Plugging these expressions into \Eref{xeq} one readily obtains the linear relation between
the strength of the forcing harmonics ($C$ and $D$) and the response of the system ($A$ and $B$) for $\nu>1$:
\beq
\!\!\!\!\!\!\!\!\!\!\!\!\!\!\!\!\!\!\!\!\!\!\!\!
\!\!\!\!\!\!\!\!\!
\left( \begin{array}{c} A_\nu \\ B_\nu \end{array}\right)
    =
    {
        \epsilon
    \over (\nu^2\omega^2/\omega_o^2-1)^2 + (\eta \nu \omega/\omega_o)^2}
        \left( \begin{array}{cc}
        1-\nu^2\omega^2/\omega_o^2 & \eta \nu \omega/\omega_o  \\
        -\eta \nu \omega/\omega_o  & 1-\nu^2 \omega^2/\omega_o^2
    \end{array}  \right)
    \left( \begin{array}{c} C_\nu \\ D_\nu  \end{array}   \right)
    \label{LinearEquation}
\eeq
The presence of lorentian resonances at $\omega_\nu = \omega_o/\nu$ is clearly visible in this
form.
As soon as a forcing exists at a certain $\nu$, the response
will be stronger near the resonance frequency $\omega_o/\nu$.
This is the reason why weak non-linearity induces structures
in the rectified current at frequencies $\omega_o/\nu$,
as shown in \Fref{fig1}.
For a static voltage there is no reference phase, and thus we can set, for instance, $B_1=0$.
The coefficients $C_1$ and $D_1$ depend through the master equation \refe{MasterEq}
strongly on $A_1$ and more weakly the other coefficients for $\nu>1$.
Equation \refe{LinearEquation} may then be solved by non-vanishing values of $A_1$ only for
a precise value of the frequency $\omega=\omega_R$.
This is very near the bare oscillation value $\omega_o$:
\beq
    \omega_R^2/\omega_o^2 \approx 1  - \eta C_1/D_1
    \label{resonanceshift}
\eeq
while the self-consistent equation for $A_1$ becomes (near the resonance):
\beq
    \eta \, A_1 = \epsilon D_1(A_1)
    \label{energybalance}
\eeq
The physical interpretation of this last equation is very simple, it represents
the energy balance between the dissipated energy  (left-hand side) and
the pumped energy (right-hand side) during one oscillation \cite{gorelik:1998}.
For small value of $a=A_1/\lambda$ it is possible to calculate analytically
$C_1$ and $D_1$, it suffices to solve the master equation by expanding
in the amplitude of oscillation $a$.
\newcommand{\tGamma}{{\tilde \Gamma}}
In our case, defining $\Gamma_L(0)=\Gamma_R(0)= \tGamma \omega_o$, at order
$a^3$ we obtain:
\beq
    C_1/\lambda  = -a {2 \tGamma \over 1 + 4 {\tGamma}^2  }
    \left[1+ {a^2 \over 4} {1-4 \tGamma^2 \over 1 +4 \tGamma^2} \right]
    \label{Ceq}
    \,,
\eeq
and
\beq
    D_1/\lambda = a { \tGamma \over 1 + 4 {\tGamma}^2  }
    \left[1+ {a^2 \over 4} {1-12 \tGamma^2 \over 1 +4 \tGamma^2} \right]
    \label{Deq}
    \,.
\eeq
The energy balance \Eref{energybalance} gives the condition
$\epsilon > \epsilon_c = \eta(1+4\tGamma^2)/\tGamma$
for the value $a=0$ to become locally unstable.
The $a^3$ term in \Eref{Deq} determines the nature of the instability.
If it is negative then $a$ increases continuously from zero when $\epsilon$ crosses the
value $\epsilon_c$.
Note also that once the problem has been solved for the principal harmonics, the
amplitude of the higher harmonics can be obtained using \Eref{LinearEquation}
the solution of the first harmonics to calculate $C_\nu$ and $D_\nu$ for
$\nu>1$.

\begin{figure}
    \centerline{
\newcommand{\larghezzadente}{5.cm}
\begin{tabular}{ccc}
   \psfig{file=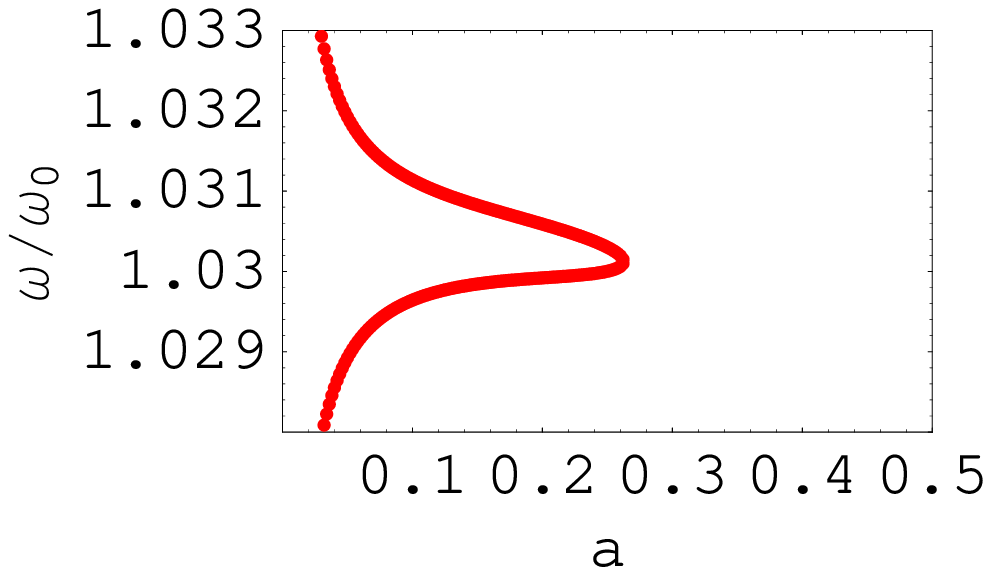,width=\larghezzadente}
    &
    \psfig{file=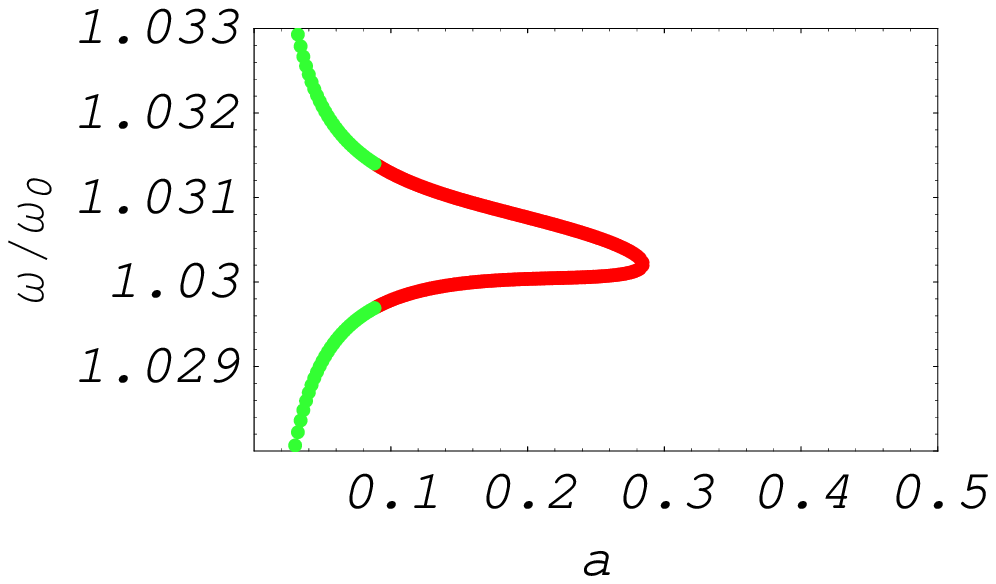,width=\larghezzadente}
    &
     \psfig{file=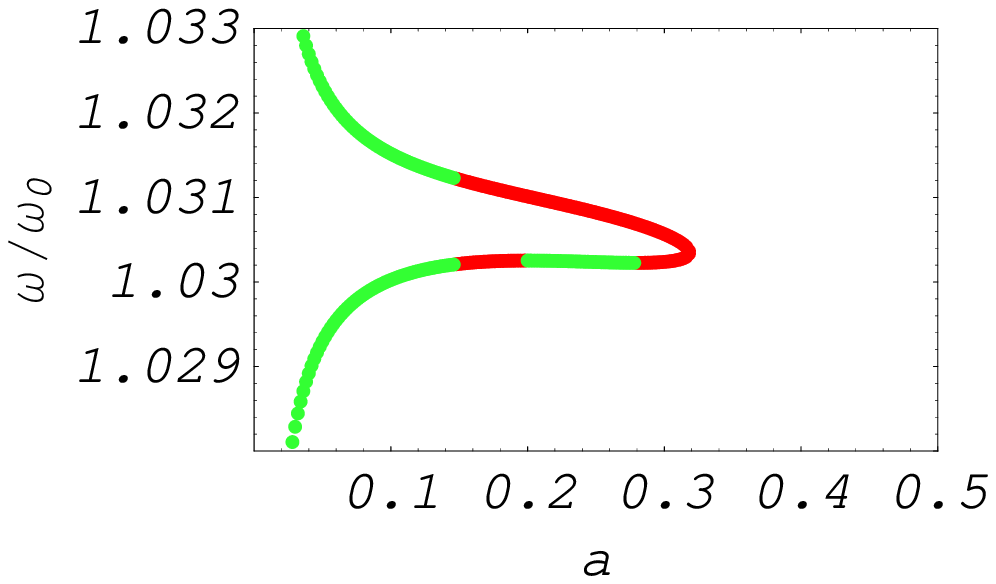,width=\larghezzadente}
    \\
   \psfig{file=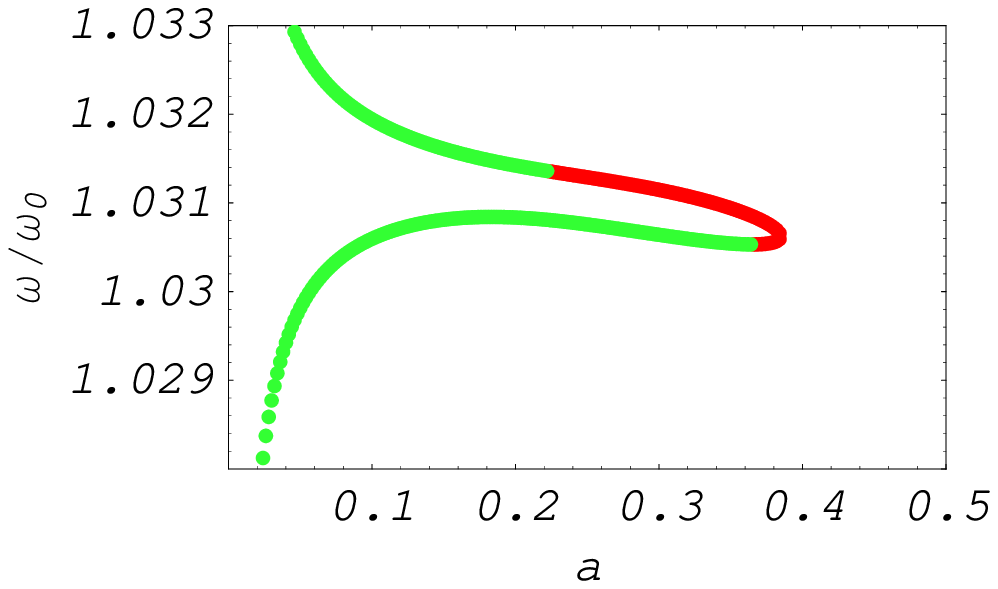,width=\larghezzadente}
    &
    \psfig{file=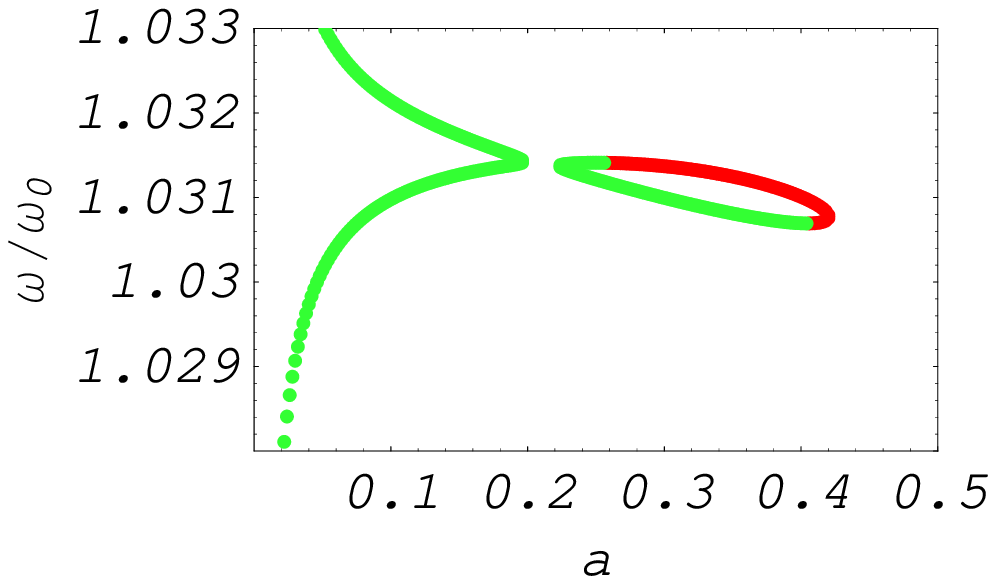,width=\larghezzadente}
    &
     \psfig{file=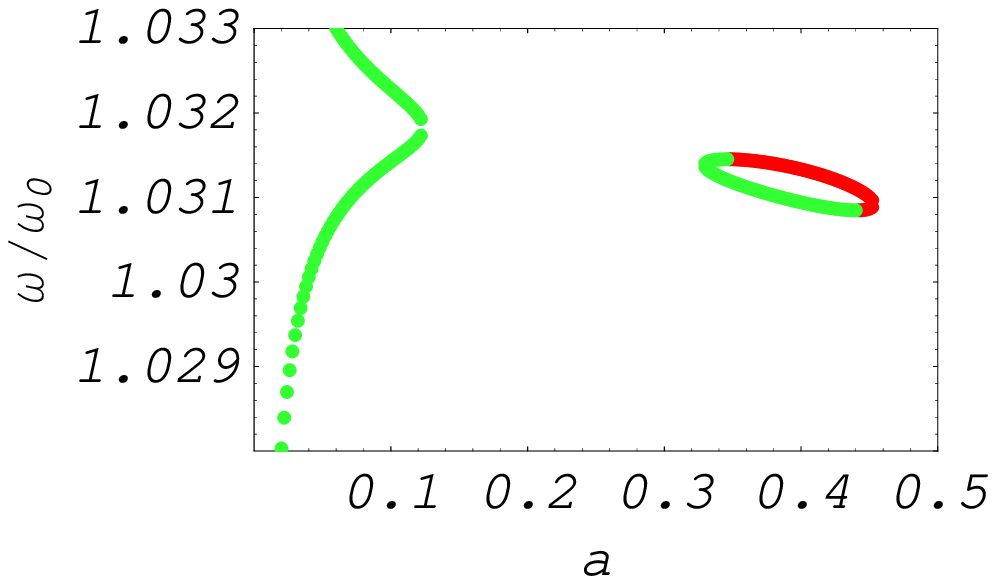,width=\larghezzadente}
\end{tabular}
}
    \caption{Evolution of the shuttle oscillation amplitude $a=A_1/\lambda$
    as a function of the forcing frequency $\omega$. The red and green points indicate
    stable and unstable values, respectively. The parameters for the plots are
    $\tGamma=1$, $\eta=0.03$, $\alpha=0.002$. This gives $\epsilon_c=5\eta$.
    Different plots differ for the ratio $\epsilon/\epsilon_c$ that takes the values 1.03, 1.035, 1.043, 1.06, 1.07, and 1.08,
    going from left to right and from top to bottom.
    The picture illustrates the evolution from a standard forced harmonic oscillator
    response to a self-oscillating shuttle.
}
\label{figdente}
\end{figure}

In the presence of a small AC additional forcing field
the equations \refe{electricforce} and \refe{LinearEquation} remain valid with
the replacement $\epsilon \rightarrow \epsilon[1+\alpha \sin(\omega t+\delta)]$ and
$P_n(t)\rightarrow P_n(t) [1+\alpha \sin(\omega t+\delta)]$, with $\alpha = V_{AC}/V_{DC} \ll 1$.
We can again set $B_1=0$, but  we need to take into account the phase shift $\delta$ of the
forcing field with respect to the oscillation.
By keeping only the principal harmonics we find that the equations for $C_1$ and $D_1$ are
modified by the appearance of the terms $\alpha \cos\delta/2$ and $\alpha \sin \delta/2$
on the right-hand side of \Eref{Ceq} and \Eref{Deq}, respectively.
The two unknown quantities $a$ and $\delta$ are determined by
Eqs. \refe{resonanceshift} and \refe{energybalance}.
For $\epsilon\ll \epsilon_c$ one finds the usual result for a damped harmonic
oscillator: $a$ has a lorentian shape and the maximum value is $\approx \alpha/[2\eta (\epsilon-\epsilon_c)]$
for $\delta = \pi/2$.
According to this expression the amplitude diverges for $\epsilon$ approaching $\epsilon_c$,
but its range of validity stops when the cubic term into \Eref{Deq} becomes important.
At that point the $\sin \delta$ term changes sign and the topological structure
of the resonance changes drastically.
In \Fref{figdente} the evolution of $\omega(a)$ as a function of $\epsilon$ is shown as obtained
from a numerical solution of the system of equations \refe{resonanceshift} and \refe{energybalance}
where $D_1$ and $C_1$ include the linear term in $\alpha$.
%

%
%
%
%
%
%

\Fref{figdente} shows  the transition from a standard harmonic oscillator
response (top-left panel),
to a self oscillating system where periodic solutions occur only for a small range of
the amplitude $a$ and the frequency $\omega$ (bottom-right panel).
The evolution is characterized by a change of the topology of the solution in the $a$-$\omega$ plane.
Increasing the static electric field ($\epsilon$) the non linearity becomes more important and at some
point (slightly larger than $\epsilon_c$) the lorentian shape is so deformed that an independent island
develops (evident in the last two plots).
In order to fully understand this evolution we need to study also the stability of the solution.
This can be done by keeping a slow time dependence of the phase $\delta$ and the amplitude $a$
in deriving the equation of motion.
Keeping only the first derivatives of $a$ and $\delta$ one obtains:
\beq
   {d \over d \omega_o t}
   \left( \begin{array}{c}  a \\ \delta \end{array}\right)
    \approx
    \left(
    \begin{array}{cc}
       1/2 & \eta/4 \\
       -\eta/4a & 1/2
    \end{array}
    \right)
    \cdot
    \left( \begin{array}{c} -\eta a + \epsilon D_1/\lambda \\ a(\omega^2/\omega_o^2-1) + \epsilon C_1/\lambda  \end{array}   \right)
    \label{timeevolution}
\eeq
Written in this way it is clear that the solution of \Eref{resonanceshift} and \Eref{energybalance} gives stationary
states (the right vector is zero) for the time evolution of the phase and the amplitude.
Studying the real part of the eigenvalues of the linear expansion of \Eref{timeevolution} around these solutions
allows the determination of their stability.
Stable solutions are indicated in red in \Fref{figdente}.

Therefore by increasing $\epsilon$ the first part of the solutions to become unstable are those
for $\omega$ far from the resonance.
We cannot predict the behavior of the system in this case, we only can say that
no periodic solution at the forcing frequency are possible.
The stable regions restricts more and more around the resonance, and the shape also
of the resonance changes drastically, as discussed above.
In the end the only stable part of the diagram is a short segment (whose size is controlled
by $\alpha$) around the solution $\omega_R, a_R$ that would be present if no AC field
was applied.

\section{Full Counting Statistics}
\label{FCS}

As we have seen the asymmetry of the resistances allows to unveil the complex
behavior of the shuttle by studying the current dependence on the forcing frequency.
The correlated charge transfer in shuttles appears to be much more visible
in the correlations of the current fluctuations.

For the shuttle, at a constant bias voltage, the noise has been considered for the first time
in \cite{weiss:1999} neglecting the correlation among different cycles.
Within a similar model, but taking into account the correlation between different cycles,
the noise and the full counting statistics (FCS) of charge transfer was then calculated by one of the
author~\cite{pistolesiFCS:2004}.
The noise taking into account the quantum nature of the oscillator was obtained by
Novotn\'y {\em et al.}~\cite{novotny:2004} by resorting to a numerical solution of
generalized Bloch equations.
An analytical approach was proposed at the same time in Ref.~\cite{isacsson:2004}.
The method to calculate the FCS in nanoelectromecanical systems proposed in \cite{pistolesiFCS:2004}
was extended to the quantum description of the oscillation in Ref.~\cite{flindt:2005}.
The FCS for a superconducting shuttle was also considered~\cite{Romito:2004}.
Knowledge of the FCS gives a deeper insight on the dynamical evolution of the nanomechanical systems.
For instance in Ref. \cite{pistolesiFCS:2004} the distribution of transmitted particles appeared to
be strongly asymmetrical, due to the fact that different parameters control the transfer of one
or two electrons per cycle.

\begin{figure}
    \centerline{
    \psfig{file=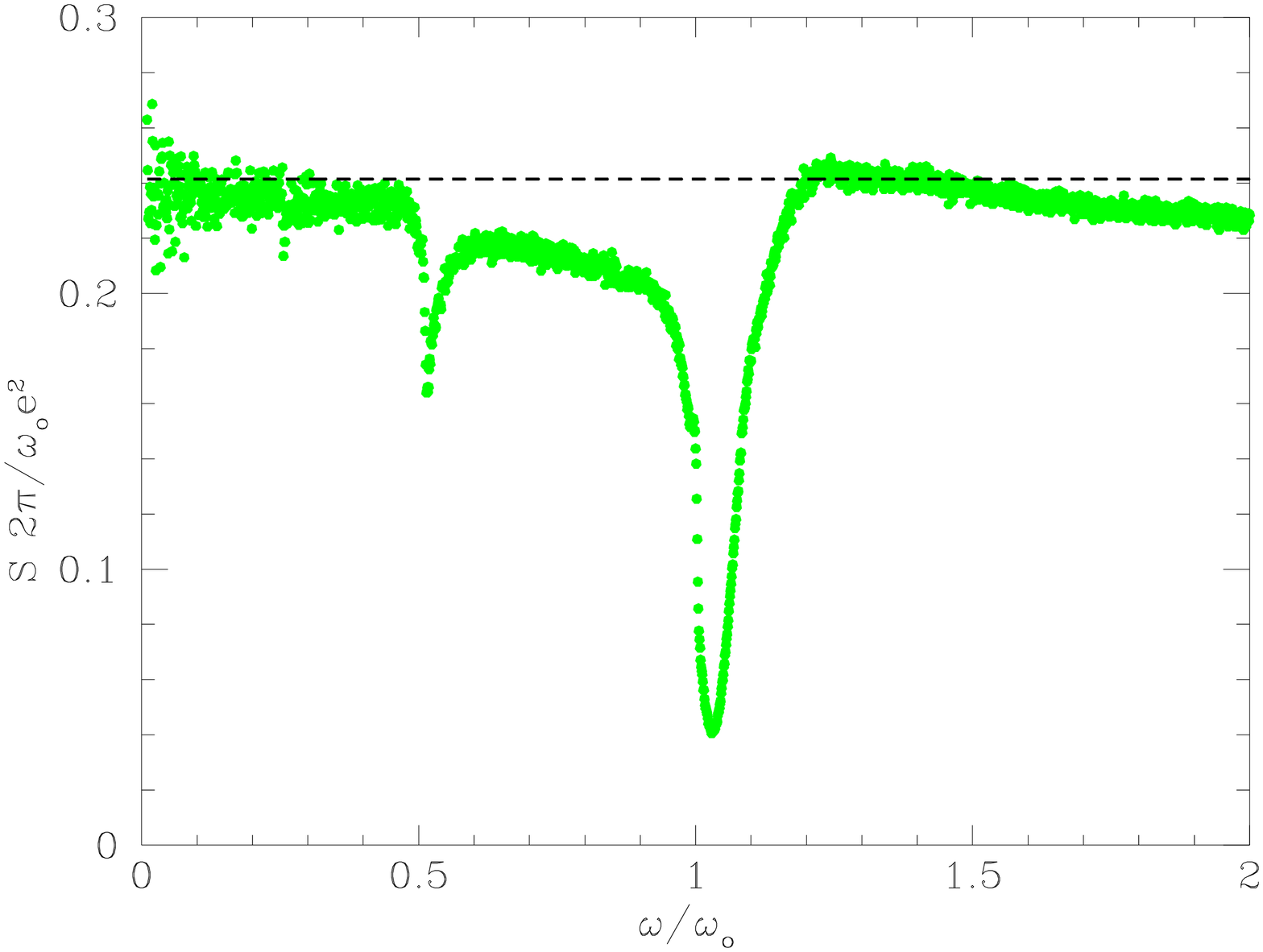,angle=-90,width=7cm}
    \hskip.5cm
    \psfig{file=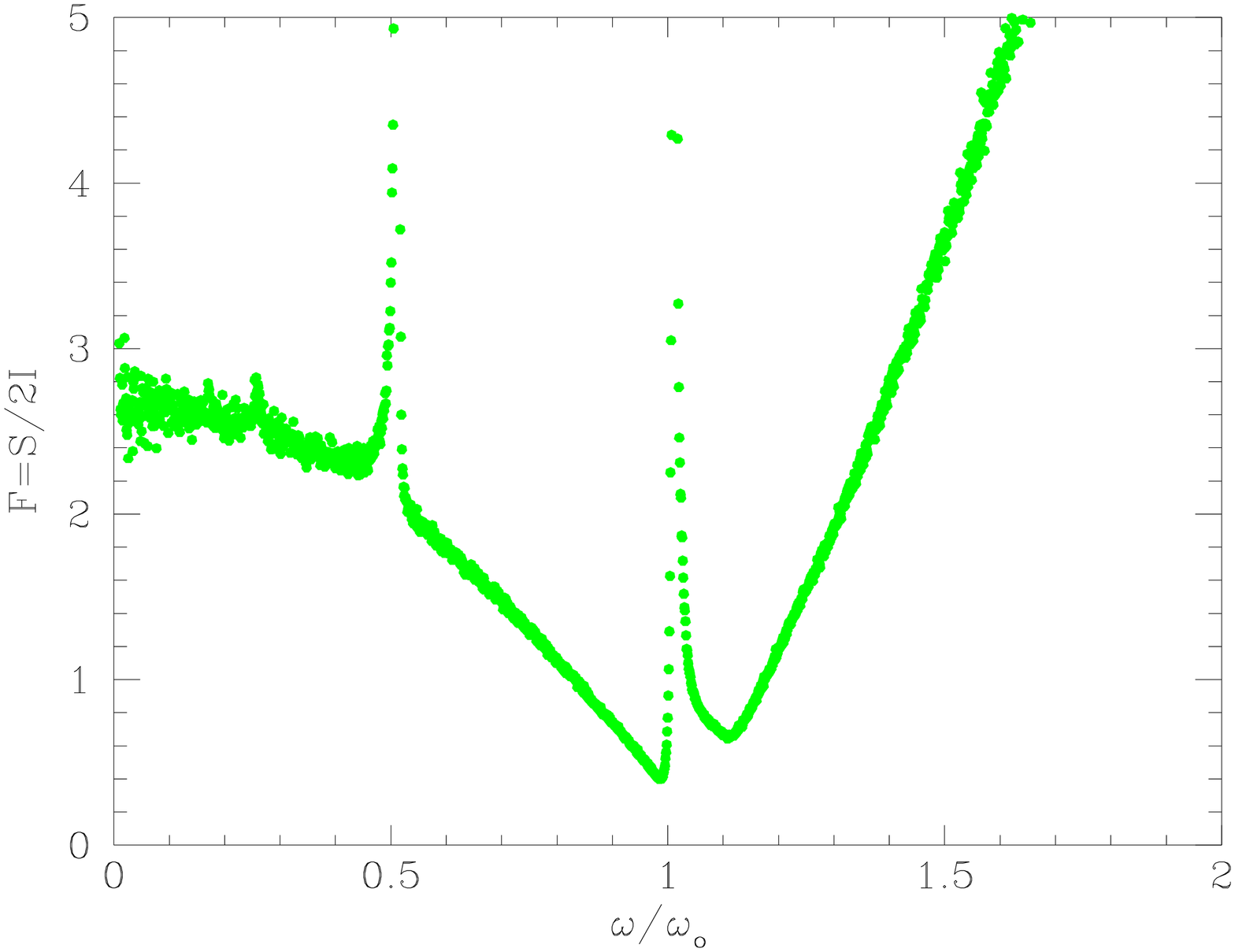,angle=-90,width=7cm}
}
    \caption{Zero frequency current noise and Fano Factor as a
    function of the external frequency. The parameters are
    the same as in \fige{fig2}). The dashed line in the left
    panel is the noise in the static SET.
}
\label{fig7}
\end{figure}

In the case of the AC-driven shuttle,  an example of the behaviour of the noise as a function of
the external frequancy is given in Fig. \ref{fig7}.
%
%
%
%
%
%
Additional interesting behaviour is expected in higher moments.
The same method discussed in Ref. \cite{pistolesiFCS:2004} can be applied to study the
statistics of current fluctuations also in presence of an AC field.

Current fluctuations can be described by the probability ${\cal P}_t(n)$ of transferring $n$
elementary charges during a measurement time $t$.
After the seminal work of Levitov, Lee and Lesovik \cite{levitov:1996} the theory to obtain the full
counting statistics has been developed and applied to many different systems
(for a review see for instance \cite{Nazarov:2002a}).
It is easier to calculate the generating function $\cal S$ that is the
logarithm of the Fourier transform of ${\cal P}_t$:
\beq
    e^{-{\cal S}_t(\chi)} = \sum _n e^{i n \chi} {\cal P}_t(n,\chi)
    \,.
    \label{generating}
\eeq
Using the definition \refe{generating} it is easy to show that the the
$n$-th derivative with respect to $i \chi$ of
$-{\cal S}$ gives the $n$-th cumulant of transmitted particles.
Explicitly the first two are the average particles $\bar n$ and the its variance
$\overline{(n-\bar n)^2}$.
For the static SET the problem has been solved by Bagrets and Nazarov \cite{bagrets:2003}.
They developed a technique to calculate in a very simple and efficient way
the FCS for an arbitrary network of contacts in the Coulomb blockade regime
(the so called "orthodox" regime, which is the one we are considering in the present  paper).
In Ref. \cite{pistolesiFCS:2004} this technique was extended to take into account the motion
of the grain, which reflects into a time dependence of the rates.
It is possible to show that the generating function can be written as follows:
\beq
    e^{{\cal S}_t(\chi)}
    =
    \langle q | T {\rm exp}
    \left\{- \int_0^t \hat \Gamma_\chi(t')\, dt'\right\} |P(0)\rangle
    \, ,
    \label{FCSNEMS}
    \eeq
here $|P(0)\rangle$ is the vector introduced in \refE{MasterEq} to describe the state of
the system, and $\langle q |= \{1,1,\dots,1\}$.
The operator $\hat \Gamma_\chi(t')$ is the rate matrix defined by \refE{Rates} where some
elements have been modified.
Explicitly if we want to count particles crossing a certain tunnel junction, we need to
attach a $e^{i\chi}$ factor to the off-diagonal elements of \refe{Rates} that induces a transfer of
charges from the leads to the grain, and a factor $e^{-i\chi}$ for the elements involving transfer of
particles in the opposite direction.
In our case if we count particles at the tunneling contact with the left reservoir we obtain:
\beq
    \hat \Gamma =
    \left(
    \begin{array}{cc}
    \Gamma_{FL}+\Gamma_{FR} & -\Gamma_{TR}-\Gamma_{TL}e^{-i\chi}
        \\
    -\Gamma_{FR}-\Gamma_{FL} e^{i\chi} & +\Gamma_{TR}+\Gamma_{TL}
    \end{array}
    \right)
    \,.
    \label{RatesChi}
\eeq
The problem is now reduced to solve the master equation \refe{MasterEq}
in presence of the counting field and coupled to the newton equation for
$x(t)$.
When the solution is periodic the expression for the generetin function can be further simplified.
In this case it is sufficient to integrate the
master equation during one period:
\beq
    \hat A = T {\rm exp} \left\{- \int_0^{2\pi/\omega} \hat \Gamma_\chi(t')\, dt'\right\}
    \,.
\eeq
The generating function for the transfer of $N$ electrons is determined by
the $N$-th power of $A$.
For large $N$ the eigenvalue $\lambda_M$ of $A$ with largest absolute value will dominate,
and the initial conditions can be neglected.
In this way the generating function becomes simply:
\beq
    -{\cal S}_N(\chi)/N = \ln \lambda_M(\chi)
\eeq
The problem is thus reduced in finding this eigenvalue.

In the following we discuss the adiabatic limit of $\omega \ll \omega_o$, $\eta \omega_o$,
and $\Gamma_L(0)+\Gamma_R(0)$.
At leading order in $\omega$ Eq.~\refe{FCSNEMS} becomes
\beq
    {\cal S}_N(\chi)/N = \int_0^{2\pi/\omega}   \lambda_1(\chi,t) dt
    \label{FCSadiabatic}
\eeq
where $\lambda_1(\chi,t)$ is the (right) eigenvalue of
$\Gamma_\chi(t)$ at time $t$ with minimum real part.
The conservation of the number of particles
implies that the real part is non negative.
Using \refE{FCSadiabatic} we calculate as an explicit example the first
four cumulants of the current fluctuation
for the AC forced shuttle.
For this purpose, we need the eigenvalue with minimal real part as a function
of time for $\hat\Gamma$ defined by \refE{Rates}.
The variable $x(t)$ is given by the solution of Newton equation neglecting the time derivatives:
\beq
    x(t) = {e V(t) \over m L } n(t) \,,
\eeq
where $n(t)$ is given by the stationary solution of \refE{MasterEq}:
\beq
    n(t) = P_1(t) = {\Gamma_{FL}(t)+\Gamma_{FR}(t) \over \Gamma_{FL}(t)+\Gamma_{FR}(t)+ \Gamma_{TL}(t)+\Gamma_{TR}(t) }
    \,.
\eeq
We consider the case $\epsilon \ll 1$, then $x/\lambda = \epsilon v(t) n(t) \ll 1$, where $v(t)=V(t)/V_o=\sin(\omega t)$.
The exponentials in $\hat \Gamma$ can be expanded.
It is convenient to separate the evolution into two parts, the range $0\leq \omega t<\pi$ with positive $v$,
and the range $\pi \leq \omega t <2 \pi$, with negative $v$.
The position satisfies at order $\epsilon$:
\beq
    {x(t)\over \lambda} = \epsilon {v\over 2}
        \left[
        (1+\beta)\, \Theta(v) + (1-\beta) \, \Theta(-v)
        \right]
        \,,
\eeq
where $\beta \equiv [\Gamma_{L}(0)-\Gamma_R(0)]/[\Gamma_L(0)+\Gamma_R(0)]$ and
we also introduce the average rate $\Gamma=[\Gamma_L(0)+\Gamma_R(0)]/2$.
The generating function at leading order in the adiabatic parameter $\gamma=eV_o\Gamma/(4E_C\omega)\gg1$ is
${\cal S}_\chi = \gamma \, \int_0^{2\pi} d\phi \lambda(\phi)$.
$\lambda(\phi)$ is the eigenvalue with minimum real part
matrix: $\hat \Gamma=\hat\Gamma_0+\epsilon \hat\Gamma_1$.
The explicit form of the matrix is:
\beqa
    \hat\Gamma_0 &=&
    |v|
    \Theta(v)
        \left(\begin{array}{cc}
            1+\beta & -(1-\beta) \\
            -(1+\beta)e^{i\chi} & 1-\beta
            \end{array}
        \right)
        \nonumber \\
    &&+|v|\Theta(-v)
        \left(\begin{array}{cc}
            1-\beta & -(1+\beta)e^{-i\chi} \\
            -(1-\beta) & 1+\beta
            \end{array}
        \right)
    \,,
\eeqa
and
\beqa
    \hat\Gamma_1 &=&
    {v |v|(1-\beta)\over 2}
    \Theta(v)
        \left(\begin{array}{cc}
            -(1+\beta) & -(1-\beta) \\
            (1+\beta)e^{i\chi} & 1-\beta
            \end{array}
        \right)
        \nonumber\\
        &&
        +
    {v |v|(1+\beta)\over 2}\Theta(-v)
        \left(\begin{array}{cc}
            1-\beta & (1+\beta)e^{-i\chi} \\
            -(1-\beta) & -(1+\beta)
            \end{array}
        \right)
        \,.
\eeqa
The eigenvalue has the form: $\lambda=\lambda_0+\epsilon\lambda_1$.
$\lambda_0$ is the eigenvalue with minimal real part of $\hat\Gamma_0$.
$\lambda_1$ can be obtained by first order perturbation theory:
$\lambda_1 = \langle f | \hat \Gamma_1 |e \rangle$, where
$|e\rangle$ and $|f\rangle$ are the right and left eigenvectors of $\Gamma_0$
corresponding to $\lambda_0$.
The calculation is straightforward, we give the result for the first four cumulants:
\beqa
    C_1 &=& \gamma \epsilon {\pi \over 4} \beta (1-\beta^2)\\
    C_2 &=& \gamma (1-\beta^2) \left[(1+\beta^2) - \epsilon {\pi\over 8} \beta^2 (1-3\beta^2)\right]\\
    C_3 &=& \gamma \epsilon {\pi \over 16} \beta (1-\beta^2) (1-12\beta^2+15 \beta^4)\\
    C_4 &=& \gamma {(1-\beta^2)\over 4}
        \left[(1+\beta^2-9\beta^4+15\beta^6)   \right.
     \nonumber \\
     && \left. -
        \epsilon {\pi\over 8} \beta^2(1-39\beta^2+135 \beta^4-105 \beta^6)
        \right]
\eeqa
$C_1$ is the average number of electrons transmitted per cycle.
It is clear from the expression that $C_1$ and all odd cumulants vanish for
symmetric structure $\beta=0$. This can be verified in general since we find
that $\lambda(\chi)$ for the first half cycle is equal to $\lambda(-\chi)$ for the
second half.
It can be interesting also to note that odd cumulants start at order $\epsilon$.
This shows that odd cumulants probe a truly nano-mechanical effect, since
$\epsilon$ parameterizes the response of the island to the electric field.
The explicit form of the first moment, the average current, is
\beq
    I_{a}(\omega \ll \omega _o)
    =
    \epsilon \, {e^2 V_o \over 32 E_C}
    {\Gamma_L \Gamma_R
        (\Gamma_L-\Gamma_R) \over (\Gamma_R+\Gamma_L)^2}
        \quad.
    \label{adiabatic}
\eeq
The corresponding value is shown in \Fref{fig2} as a dashed line.
The (small) difference with the numerical result comes from the $\epsilon$ expansion,
($\epsilon=0.5$ in the simulation) as is clear with a comparison with the black triangle
obtained by solving numerically the adiabatic equations for the current.

\section{Conclusions}
\label{Conclusions}

In this work we analyzed the properties of the charge shuttle under time-dependent
driving showing that the response of the shuttle is rather rich due to the non-linear
dynamics of the grain.
Rectification, synchronization, response to multiple of the external frequency appear
in the electrical properties of the shuttle.
All the results obtained here can be tested experimentally.
Indeed a recent paper by Scheible and Blick~\cite{scheible:2004} already reports on
some of the properties which we saw in the average current.
Part of these results were presented elsewhere~\cite{pistolesiFCS:2004,pistolesi:2005}.
Here we provided a new description on the dynamical behavior of the system,
the transition from the stationary to the self-oscillating solution, and
the Full Counting Statistics in the presence of an AC-driving.

\section*{Bibliography}

\bibliography{biblioNEMS,biblio}

\end{document}